\documentclass[12pt]{amsart}
   	
\usepackage{graphicx}
\usepackage{amssymb}
\usepackage{booktabs}
\usepackage{multirow}
\usepackage{hhline}
\usepackage{fancyhdr}

\newcommand{\fr}[2]{\frac{#1}{#2}}

\newcommand{\ds}[0]{\displaystyle}
\newcommand{\pow}[1]{\times10^{#1}}

\title[The implied longevity curve]{The implied longevity curve: How long does the market think you are going to live?}

\author{Moshe A. Milevsky, Thomas S. Salisbury, and Alexander Chigodaev}
\thanks{Alexander Chigodaev was a post-doctoral fellow in the Department of Mathematics and Statistics, York University, when this work was conducted. Milevsky (the contact author) is an Associate Professor of Finance at the Schulich School of Business, York University, and Executive Director of the IFID Centre. He can be reached at milevsky@yorku.ca ; Salisbury is a Professor in the Department of Mathematics and Statistics at York University.  The authors acknowledge funding from the IFID Centre (Milevsky) and from NSERC (Salisbury) and thank Simon Dabrowski (CANNEX) for assistance. This work was made possible by the Shared Hierarchical Academic Research Computing Network (SHARCNET:www.sharcnet.ca) and Compute/Calcul Canada. Disclosure Note: Two authors (Salisbury and Milevsky) have a financial relationship with CANNEX, the source of the data. An earlier version of this paper circulated under the title {\it How Long Does the Market think You Will Live? Implying Longevity from Annuity Prices.}} 
			
\date{15 June 2014 (Version 2.1)}				

\begin{document}

\begin{abstract}
We use life annuity prices to extract information about human longevity using a framework that links the term structure of mortality and interest rates. We \emph{invert the model} and perform nonlinear least squares to obtain implied longevity forecasts. Methodologically, we assume a Cox-Ingersoll-Ross (CIR) model for the underlying yield curve, and for mortality, a Gompertz-Makeham (GM) law that varies with the year of annuity purchase. Our main result is that over the last decade markets implied an improvement in longevity of of 6--7 weeks per year for males and 1--3 weeks for females. In the year 2004 market prices implied a $40.1\%$ probability of survival to the age $90$ for a $75$-year old male ($51.2\%$ for a female) annuitant. By the year 2013 the implied survival probability had increased to $46.1\%$ (and $53.1\%$). The corresponding implied life expectancy has increased (at the age of 75) from $13.09$ years for males ($15.08$ years for females) to $14.28$ years (and $15.61$ years.) Although these values are implied directly from markets, they are consistent with demographic projections. Similar to implied volatility in option pricing, we believe that our implied survival probabilities (ISP) and implied life expectancy (ILE) are relevant for the financial management of assets post-retirement and very important for the optimal timing and allocation to annuities; procrastinators are swimming against an uncertain but rather strong longevity trend.

\end{abstract}
\maketitle

\newpage

\section{Introduction}
One of the corollaries of the efficient market hypothesis (EMH) is that market prices contain valuable information and provide a consensus view about the discounted economic value of the future. And, while this position is controversial these days (in light of evidence from behavioral finance), market prices continue to be used to extract information. Indeed, futures, options and various derivatives are often used to forecast stocks, commodities, interest rates, volatility and even the weather. So, motivated by the paradigm around the use of prices to extract information about the future, in this paper we use market prices of life annuities to imply information about expected human longevity and its improvements over time. 

In the past this was hard to achieve, as it was difficult to obtain a reliable and consistent cross-sectional time series of life annuity or insurance prices.\footnote{See the working paper by Milevsky, Jiang and Promislow (2001) for an early attempt.} Now however, we have access to a data set consisting of over 2 million U.S. annuity quotes sampled at weekly frequencies over a period of a decade. This unprecedented data together with a powerful computational procedure -- both of which are described in the body of the paper -- enable us to translate prices into longevity expectations in a way that has not been done before.

\begin{figure}[!h]
  \center{\includegraphics[scale=0.35]{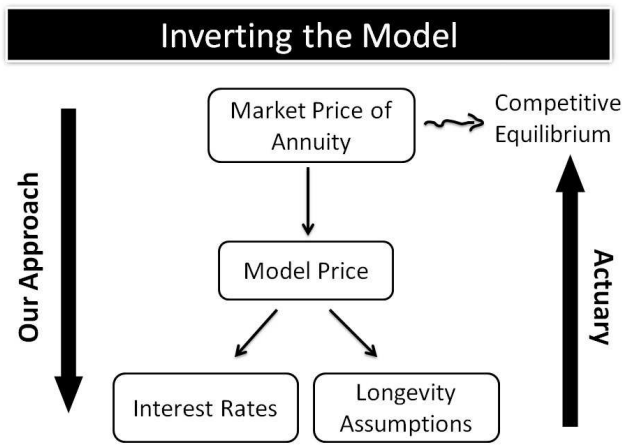}} 
  \caption{Normally, insurance actuaries start with (i.) longevity expectations which are embedded in demographic mortality tables together with (ii.) interest rate assumptions. Consequently, they solve for annuity model prices which then lead to market prices. We invert the process and use market prices observed in a competitive equilibrium to solve for implied longevity expectations and to gauge how they change over time.}
  \label{FIG:INVERTMODEL} 
\end{figure}

Technically speaking, we invert the model -- which links the term structure of mortality and interest rates -- and perform nonlinear least squares fitting to obtain implied mortality parameters. The inversion process is displayed graphically in Figure (\ref{FIG:INVERTMODEL}) {\bf [placed here]}. Recall that normally the insurance actuary begins with a set of assumptions about future interest rates and mortality. He/she then computes a model annuity price, which then becomes the market price after adjusting for competitive factors. In this paper we reverse the process. We start with the market price, which we assume is an amalgamation of a consensus view of longevity, then we solve for the implied tables and parameters. This is quite similar to the procedure used by Finkelstein and Poterba (2004), or Mullin and Philipson (1997) in which they used insurance policy and annuity prices to extract information about adverse-selection and mortality expectations, although our procedure is more dynamic and the data more refined. More on this later.

Implied survival probabilities and life expectancy are relevant for the financial management of assets post-retirement as well as the optimal timing and allocation to annuities. Our main result is that over the last decade (for which we have reliable prices), life annuities imply a significant improvement in longevity. Thus, while in the year 2004 market prices implied a 40.1\% probability of survival to the age 90 for a 75-year old male (51.2\% for a female) annuitant, by the year 2013 the implied survival probability (ISP) had increased to 46.1\% (and 53.1\%). In conjunction, we find that the corresponding implied life expectancy (ILE) has increased (at the age of 75) from 13.09 years for males (15.08 years for females) to 14.28 years (and 15.61 years) over the same period. This corresponds to 6.8 weeks per year for males (3.0 weeks for females). These numbers are roughly consistent with demographic and actuarial projections for the improvement in longevity -- for example the widely publicized work by James Vaupel at the Max Planck Institute\footnote{See Oeppen and Vaupel (2002) as well as Lee and Carter (1992) for the statistical or biological approach to projecting longevity and the estimates provided for mortality improvements.} -- but are obtained from market prices directly.

\subsection{Agenda and Plan}

The remainder of this paper is organized as follows. Section \#\ref{SECprice} describes the pricing of life annuities. Section \#\ref{SECgompertz} explains the underlying law of morality. Section \#\ref{SECdata} describes the source and structure of our annuity price data set. We describe our numerical results in section \#\ref{SECresults} and draw conclusions in section \#\ref{SECconclusions}. We relegate all technical model details to the appendix. 

\section{Annuity Prices}
\label{SECprice}

The price of a life annuity -- or inversely, the amount of income a retiree can expect for a given premium deposit -- is determined in a competitive market based on the interaction of numerous insurance companies. Even so, while the final price paid is partially determined by the forces of supply and demand, there is a strict mathematical relationship linking mortality expectations and interest rates to observed prices. This is akin to the concept of arbitrage in securities markets, where the market prices can't deviate too much from certain model values. 

The simplest pricing formula for a life annuity that pays \$1 each year for life is as follows:
\begin{equation}
  a(x, T, R) = \sum_{i=1}^{T} \fr{1}{(1+R)^i} + \sum^{\omega-x}_{i=T+1} \fr{p(x, i)}{(1+R)^i}.
  \label{EQ:ANTDISC} 
\end{equation}
This assumes that payments take place once per year, and that the term structure is flat, neither of which will actually be true in our final model. Nonetheless this basic version will serve to illustrate the idea.

The quantity $a(x, T, R)$ denotes the up-front ``cost'' of \$1 per year for life, starting at the age of $x$, guaranteed for $T$ years. By the guarantee period, we mean that if the annuitant dies during this period, payments continue to a designated beneficiary until the end of the period. On the right-hand side there are two sums: the guaranteed portion and the life-contingent portion. In the life-contingent portion, the ratio of the survival probability $p(x, i)$ and the interest rate factor $(1+R)^i$ are added-up until the end of the mortality table. The sum terminates at the age of $\omega$, where ``omega'' denotes the oldest possible age attainable, currently 122 years\footnote{Jeanne Louise Calment, a French supercentenarian has the longest confirmed human lifespan in history, living to the age of 122 years, 164 days.}.

Equation (\ref{EQ:ANTDISC}) differs from the standard present value (PV) formula familiar to financial analysts and wealth managers by having a survival-contingent probability instead of the standard \$1 in the numerator of the second summation. Think of this equation as the present value factor of one dollar of income to be received for as long as you are alive. For example, if you are 70 years old and the probability of surviving for 1 year is 97\%, for 2 years is 95\% and for 3 years is 92\%, then the first three terms of the life contingent present value embedded in equation (\ref{EQ:ANTDISC}) are: $$\ds\fr{0.97}{(1.05)}+\fr{0.95}{(1.05)^2}+\fr{0.92}{(1.05)^2}$$ with the remaining terms declining in importance until the final numerator is effectively zero. 

One could use any (declining) survival probability vector of numbers in the numerator of equation (\ref{EQ:ANTDISC}), and add-up the terms to arrive at $a(x, T, R)$. However, for the purposes of this study and this paper, we assumed a particular functional form, which is known as the Gompertz-Makeham (GM) law of mortality. This functional form is quite common in the biological and demographic fields, and is used in economics as well. Alas, without imposing some sort of structure on mortality, it would be nearly impossible to extract expectations.

\section{Gompertz-Makeham Law Explained}
\label{SECgompertz}

Actuaries and demographers have long-established that age-dependent single-year mortality rates increase with age by approximately 8\% to as high as 10\% every year between the ages of 25 and as possibly as high as the age of 95. In other words, if someone at age $y$ has a $q$ percent probability of dying within that year, and mortality increases by 8\% annually, then if they survive to age $(y+1)$ their probability of dying within the next year becomes $q(1+0.08)$ percent, then $q(1+0.08)^2$ percent at age $(y+2)$, and then $q(1+0.08)^3$ percent at age $(y+3)$, etc. To a first order of approximation, human mortality rates (for adults), regardless of what particular mortality table you select, are an exponentially increasing function of age. Therefore, by computing the logarithms of the annual mortality rates, they can be approximated by a straight line and characterized by a slope and an intercept.

This biological observation was first made by the British demographer and actuary Benjamin Gompertz (b. 1779, d. 1865) and later refined by William Makeham in 1890, and is today known as the Gompertz-Makeham (GM) “law of mortality” in their honour.  The Gompertz portion of the GM law refers to the exponential increase mentioned above, and the Makeham portion refers to a constant rate of accidents that is independent of age. For more information about these laws of mortality and the actuarial minutia behind these calculations, see the standard actuarial texts by Promislow (2011) or Dickson, Hardy and Waters (2009).

The GM formulation gives a powerful analytic tool to compute survival probabilities to any age as a function of just three basic parameters.  The concise expression for the survival probability -- from any age, to any time -- under the GM law of mortality can be written as follows:
\begin{equation}
  \ln p(x, t) = - \lambda_0 \, t + (1 - e^{t/b}) \, e^{(x-m)/b}
  \label{EQ:GOMAKLAW} 
\end{equation}
where the variable $t$ denotes the survival time, the variable $x$ denotes the current age of the individual, $\lambda_0$ denotes an accidental rate of death and the parameters $(m,b)$ denote the modal value and the dispersion coefficient, both measured in units of time. The survival probability itself -- and the main quantity of interest -- is trivially obtained by taking the exponent of the right-hand side of equation (\ref{EQ:GOMAKLAW}). 

Given the centrality of the GM law to our ``implied longevity'' algorithm, we now provide a detailed example of how to use the parameters to arrive at numerical values. Assume that you are currently 50-years old and would like to estimate the probability you will live (at least) to the age of 90, which is 40 more years. According to the GM law of mortality, this probability depends on three parameters; the accidental death rate $\lambda_0$, the modal value of life $m$ and the dispersion value of life $b$. The last two numbers can loosely be thought of as playing the role of the mean and standard deviation of your remaining future lifetime when $\lambda_0$ is zero. Technically speaking, the modal value of life is the age at which an individual is most likely to die\footnote{This assumes they are younger than $m$ to start with.}, but is actually a few years higher than the 50/50 (median lifespan) point. This is due to the skewness of the distribution. Technicalities aside, if -- for example -- the modal value is $m=80$ years and the dispersion value is $b=11$ years (assuming $\lambda_0$ is zero), then according to equation (\ref{EQ:GOMAKLAW}), the survival probability to age 90 is 8.9\%, which can also be expressed as a 91.1\% probability of dying prior to age 90. In contrast, (still assuming $\lambda_0$ is zero) under a higher modal value of $m=92$ years in equation (\ref{EQ:GOMAKLAW}), instead of the lower $m=80$, the survival probability to age 90 increases to 44.4\%. Note how the extra 12 years of life (in the modal sense) will add 35.5 percentage points to the survival probability. In fact, if you ``believe'' that your modal value of life is in fact $m=92$ years, then according to equation (\ref{EQ:GOMAKLAW}) the probability of surviving to age 95 (from age 50) is 27.5\% and the probability of surviving to age 100 and becoming a centenarian is 12.9\%. 

It is an empirical observation that the mortality rates given by the GM model are a very good approximation to observed mortality rates, when it comes to pricing life annuities at retirement ages 55 to 80. Instead of using equation (\ref{EQ:ANTDISC}), in which the annuity pays only once per year, we consider an annuity paying \emph{continuously} in time. In that case it is possible to obtain a closed-form expression for the life annuity factor when mortality is assumed to obey the Gompertz-Makeham (GM) law, and the interest rate is constant. The expression involves the incomplete Gamma function $\Gamma(a,b)=\int_b^\infty e^{-s} \, s^{a-1} \, ds.$ which is found in almost all mathematical software packages. We refer the interested reader to Milevsky (2006, 2012) for more information. The Gompertz Annuity Pricing Model (GAPM) for a $T$ years period certain life annuity is then:
\begin{equation}
  \bar{a}(x,T,r) = \frac{1-e^{-rT}}{r} + \fr{b \, \Gamma\left( -b \left( r + \lambda_0\right), \exp\left\{\fr{x - m + T}{b}\right\}\right)}{\exp\left( (r + \lambda_0) \, (m - x) - \exp\left\{\fr{x-m}{b}\right\}\right)}
  \label{GAPMformula}
\end{equation}
where the letter $r$ denotes the continuously compounded interest rate, or $r = \ln(1+R)$. The usage of the bar on top of the $\bar{a}$ represents continuous time payments. If interest rates and mortality parameters remained constant, then annuity prices would not change over time, and would be captured precisely by the GAPM formula \eqref{GAPMformula}. In practice, prices and parameters do change, so we will modify the GAPM model in a manner that is described in the technical appendix.

To conclude, while no formula in finance (even Black Scholes) can provide a perfect fit to the observed price of a financial asset traded in the market, the ``life annuity factor'' described by equation (\ref{EQ:ANTDISC}) using the mortality rates from equation (\ref{EQ:GOMAKLAW}) provide a reasonably good fit to quotes offered by insurance companies. Our objective then is to use market prices of annuities to imply these parameters and how they change over time. 

\section{Source of Data}
\label{SECdata}

The raw data used in our analysis was obtained using a three-step process. In the first step we utilized the ``survey'' services of CANNEX Financial Exchanges, which is a data and analytics vendor with an ongoing business relationship with most U.S. insurance companies that sell and market life annuities. CANNEX has the ability to dial into the insurance company's internal servers and extract any pre-specified life annuity quote for a wide range of parameters, such as, age, gender, period certain, etc. Financial advisers, insurance brokers and wealth managers utilize CANNEX's services to obtain current quotes (which in some cases are binding on the insurer). They are the most widely used and comprehensive source for up-to-date life annuity quotes. CANNEX charges a minimal fee per ``survey'' to subscribers to their system which is how they generate revenue. In the U.S. annuities can be purchased with after-tax (qualified) funds or with pre-tax (non-qualified) funds. Although both types of quotes are available from the CANNEX system, we decided to utilize the non-qualified (NQ) annuity data for a variety of data-integrity and reliability issues. Keep in mind that part of the NQ annuity income is tax-free (return of principal) and part is taxable. 

From the insurance company's perspective, these quotes are the prices at which they would be obligated to offer that particular (NQ) life annuity. This is analogous to calling a foreign exchange dealer and asking them their bid or ask price on given quantity of a particular currency for a specific delivery date. The CANNEX system effectively samples all insurance companies at the exact time enquiring about the price of many different annuities. We stress that insurance companies are obligated to honor those quotes for a few days, which makes this more than a hypothetical query.

In the second stage of the data generating process, CANNEX would send a data file of many age/guarantee quotes to QWeMA Group (a software company) on a weekly basis, which QWeMA Group would check and validate for internal consistency and then store in a (large) database.  This database now consists of over 2 million quotes. The QWeMA Group has constructed various indices for annuity prices based on the averages of these quotes. The company has made summaries and various time-series available to researchers\footnote{Note. Recently, QWeMA was purchased by CANNEX}.

Finally, we (the authors) obtained weekly data from the QWeMA Group for a period of 478 weeks starting Sept 15, 2004 (including 9 weeks for which data was missing). The data consists of an average annuity quote (averaged over companies, trimming the highest and lowest) for each of 7 ages (50--80 years), 6 guarantee periods (0--25 years), and 2 genders. Henceforth, we refer to this as the CANNEX-QWeMA data set and gratefully acknowledge both company's employees for assistance in this research.

\section{Analysis of Results}
\label{SECresults}

The annuities quoted provide monthly payments, which are well approximated by continuous payments, 
as in \eqref{GAPMformula}. For quotes issued at week $z$, we assume that pricing is based on a spot interest rate $r_z$ and a choice of Gompertz-Makeham (GM) parameters as displayed in equation (\ref{EQ:GOMAKLAW}), which we label $\lambda_0$, $b(z)$, and $m(z)$. As described in the appendix, we assume $b(z)$ and $m(z)$ are linear in $z$, so there are five GM parameters in total. We invert the modified/enhanced versions of the resulting equations and solve for the various spot rates and GM parameters, which leads to ISPs and ILEs on a weekly basis.

We emphasize that we are \emph{implying} weekly spot interest rates from the annuity quotes. In theory, we could have observed these in the market -- for example the U.S. Treasury bond curve, or the Corporate bond curve. But in fact, it is unclear which particular exogenous spot curve should be used in the context of annuities (and since our data are averaged over companies, that may set the spot rate in different ways, it is plausible that NO exogenous curve is exactly ``right'').  In keeping with the approach of this paper, that we let market prices ``tell'' us about parameters, we do this for spot rates as well. See Figure (\ref{FIG:INVERTMODEL}).

The one last modelling item we need to address is how the term structure of interest rates factors into the annuity pricing equation. In other words, we need to specify the assumptions the annuity provider will make about the future evolution of interest rates. We carried this out in two different ways; a simple model and a realistic model. For the simple model, we assumed providers work with a flat term structure, so that once that week's spot parameter is estimated, formula \eqref{GAPMformula} applies. We call this the \emph{flat model}. This was a first step to give some intuition for changes in implied longevity. Then, we implemented a more sophisticated model (which we called the \emph{curved model}) in which we assumed that providers adopt a functional form known as the Cox-Ingersoll-Ross (CIR) (1985) model. This adds three CIR parameters (assumed to be constant across the 10 year period) which we also imply from the data, along with the implied longevity parameters, and the weekly short rate values (as latent variables). We present both sets of results, as this gives a sense of the results' sensitivity to the term structure model. In the technical appendix, we provide additional details on precisely how the inversion procedure was implemented. We now move-on to describe the results.

Table (\ref{TAB:PARSFLAT}) {\bf [placed here]} provides a summary of the parameters under our simple (flat) model. When the valuation term structure is assumed to be flat, there are a total of five free parameters to be estimated in the (simple) model. They are $\lambda_0$, which is the accidental death rate, the initial mode and scale parameters ($m_0,b_0$) respectively, and the rates at which they improve over time ($m_1,b_1$). The table displays their point estimates. So, for example, in the simple (flat) model the implied modal value of life (in 2004) was 93 years for females and 88 years for males. Note that even our simple yield curve model indicates that mortality improves over time. 

Table (\ref{TAB:FTEST}) {\bf [placed here]} provides a summary of the partial $F$-statistic test in the simple (flat term structure) model. $F$ is intended to measure the reduced error dispersion when using a mortality model with improvements. It is a test of whether mortality expectations are changing over time. The expression $F^{\left[p - g, n - k \right]}_{\left[\beta \right]}$ are the 1-$\beta$th percentiles of the $F$-distribution. What all of this means is that (even in the simple model) we can reject the null hypothesis that $m_1=0,b_1=0$ and that there are no changes in implied mortality over time. In practical terms this means that -- for instance --  a 65 year-old in 2004 is quite different from a 65 year-old in 2013 from a life expectancy point of view.  

\begin{figure}[!h]
  \center{\includegraphics[scale=0.65]{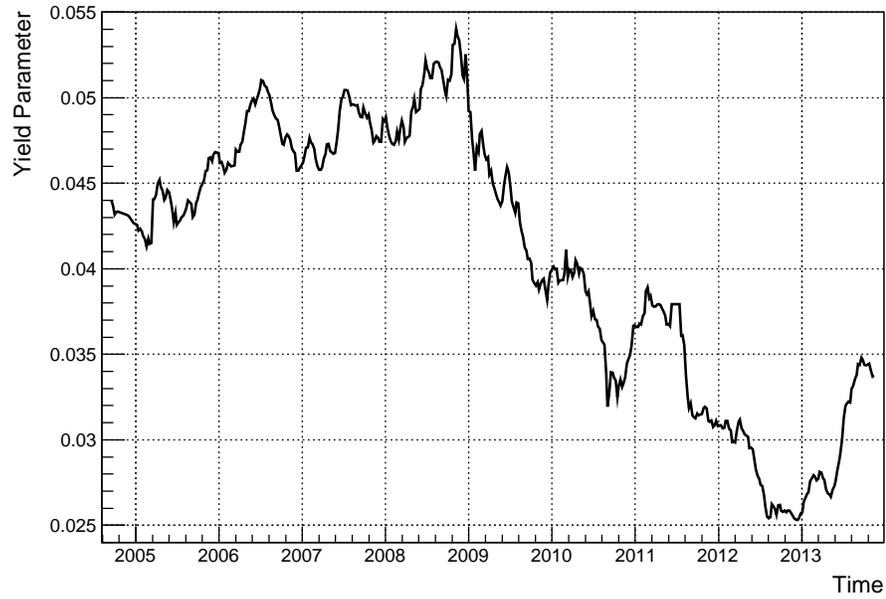}} 
  \caption{Estimated values for the yield (spot rate) parameter $r_z$ in the flat term structure model.} 
  \label{FIG:FFLAT} 
\end{figure}

In Figure (\ref{FIG:FFLAT}) {\bf [placed here]}, we display the estimated (weekly) values for the yield $r_z$ (i.e. spot rate parameter) in the simple (flat term structure) model. These numbers should be interpreted as the summary or average interest rate that insurance companies (in aggregate) were using to discount cash-flows for pricing life annuities. During the 2004 to 2013 period the highest rate was 5.5\% (around October 2008) and the lowest rate was 2.5\% (second half of 2012).

In the the more realistic (curved) model where we assumed a CIR term structure underlying the pricing of annuities, we estimated three free interest rate (diffusion) parameters in addition to the instantaneous (spot) rate $r_z$. These are displayed in Table (\ref{TAB:PARSCIR}) {\bf [placed here]}. The difference in parameters between males and females does not necessarily imply that one is getting a better rate than the other. Rather, this is a reflection of the presence of numerical noise and that the model does not perfectly distinguish between mortality and interest rate parameters.

Table (\ref{TAB:PARSMORT}) {\bf [placed here]} displays the mortality parameter estimates in the curved model. These values should be compared against the implied mortality parameters in the flat model displayed in Table (\ref{TAB:PARSFLAT}). Notice that once a more realistic model of the yield curve is included in the pricing process, the (modal) rate of mortality improvement becomes higher for males than for females. This is consistent with demographic data, but could also be a reflection of the difficulty in disentangling interest rates from mortality.

\begin{figure}[!h]
  \center{\includegraphics[scale=0.65]{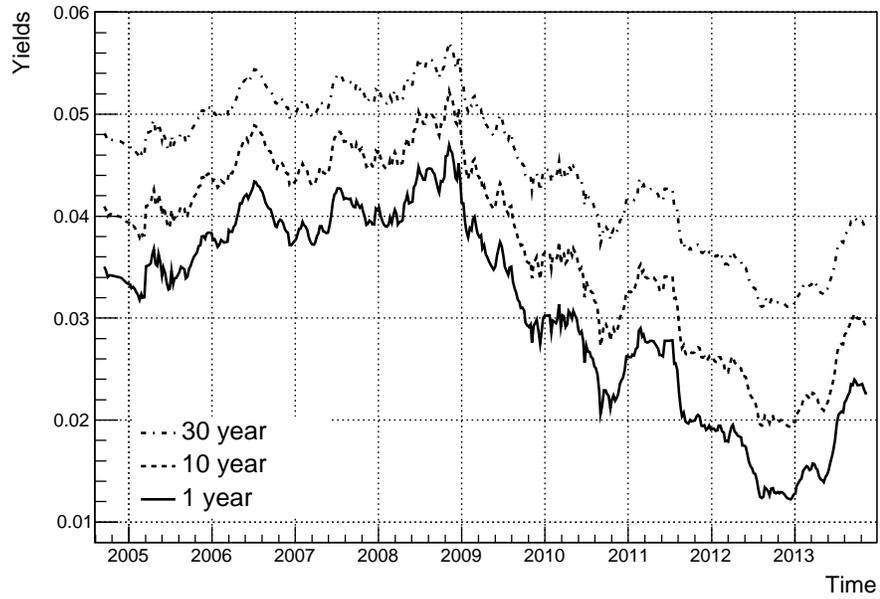}} %
  \caption{Yields of varying maturities, implied by the CIR (curved) term structure model for each week in the sample.}
  \label{FIG:CIRYIELDS} 
\end{figure}

Figure (\ref{FIG:CIRYIELDS}) {\bf [placed here]} displays the implied interest rates at various points on the yield curve for the duration of the considered time period. Recall that in the CIR model underlying our estimates, for each week in the sample, the instantaneous spot interest rate $r_z$ together with the three global parameters $\alpha,\beta,\sigma$ will induce an entire yield curve. In this figure, we display the rate for maturities of 1 year, 10 years and 30 years. In some sense, Figure (\ref{FIG:FFLAT}) should be viewed as a weighted average of these numbers. 

\begin{figure}[!h]
  \center{\includegraphics[scale=0.65]{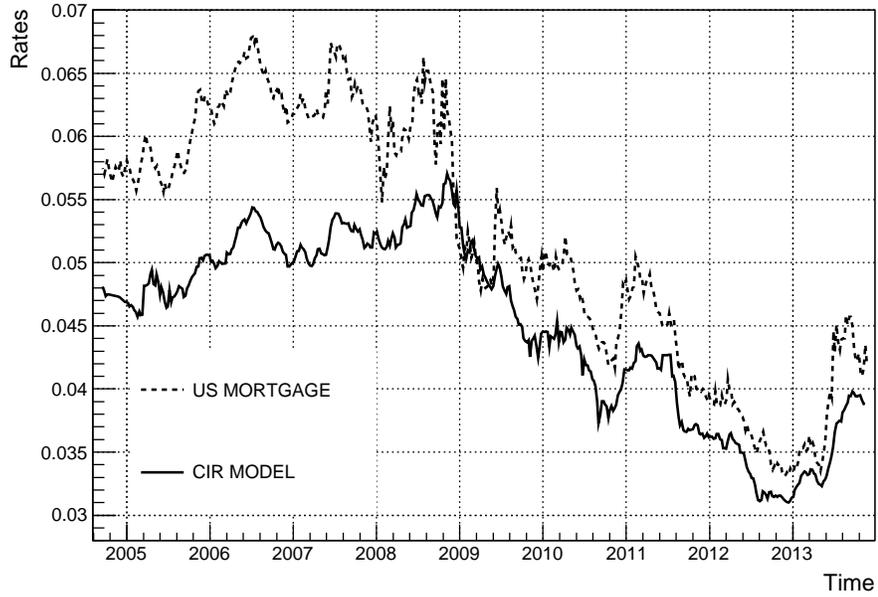}} %
  \caption{Comparison between 30 year yields implied by the CIR (curved) model, and 30 year US mortgage rates.}
  \label{FIG:CIRMORT} 
\end{figure}

As a means of comparison, Figure (\ref{FIG:CIRMORT}) {\bf [placed here]} superimposes the 30 year U.S. mortgage rate (based on Federal Reserve Data) and the 30 year interest rate implied from the CIR model over the period we analyzed. Based on discussions with insurance company actuaries, the 30 year mortgage rate is an important determinant of annuity prices. Generally speaking, they track each other well, although the spread seems to have declined after the financial crisis.

\begin{figure}
  \center{\includegraphics[scale=0.65]{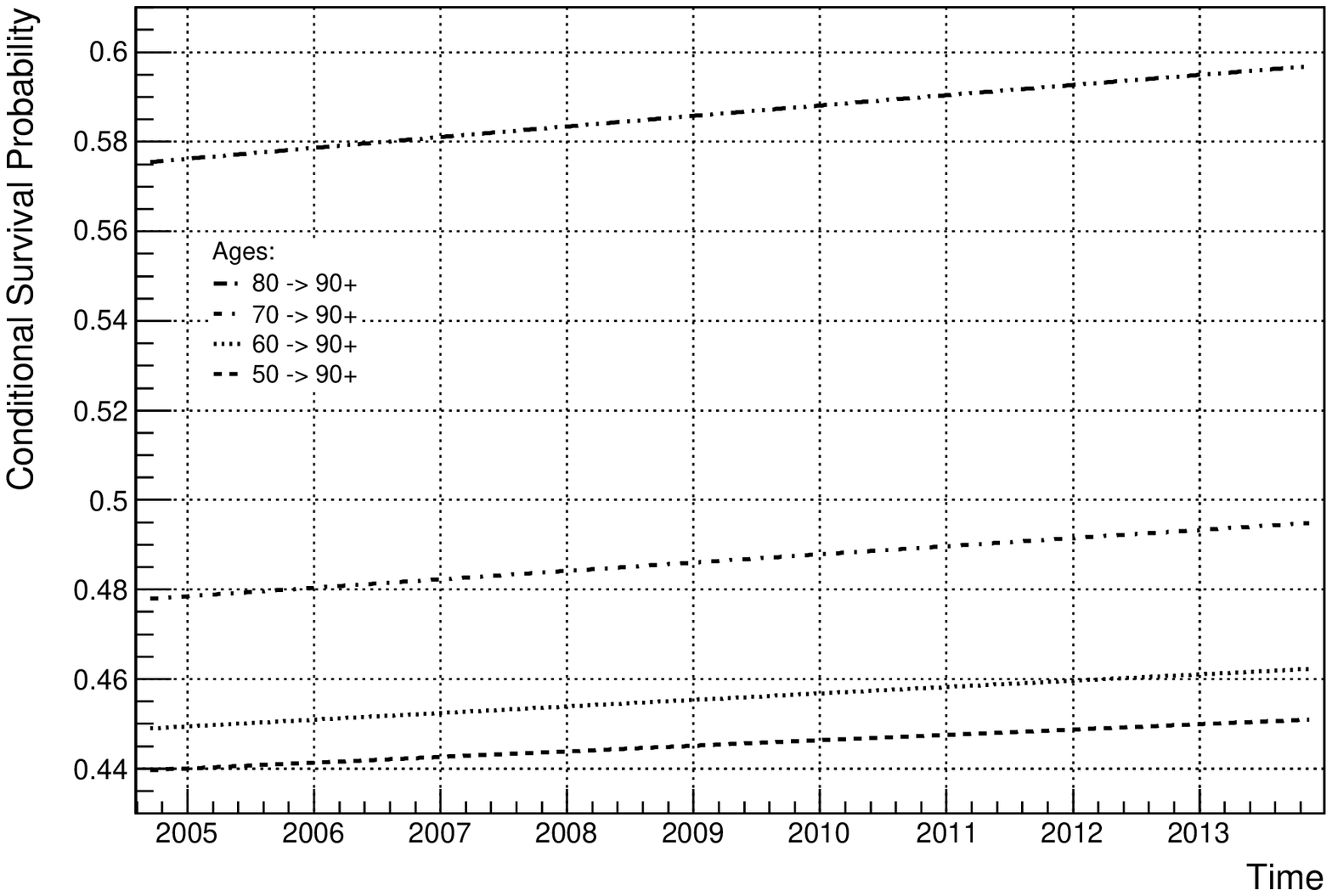}} 
  \caption{Implied Survival Probabilities (ISPs) derived from the GM parameters of the (curved) CIR model for FEMALES.}
  \label{FIG:FSBCIR} 
\end{figure}

\begin{figure}
  \center{\includegraphics[scale=0.65]{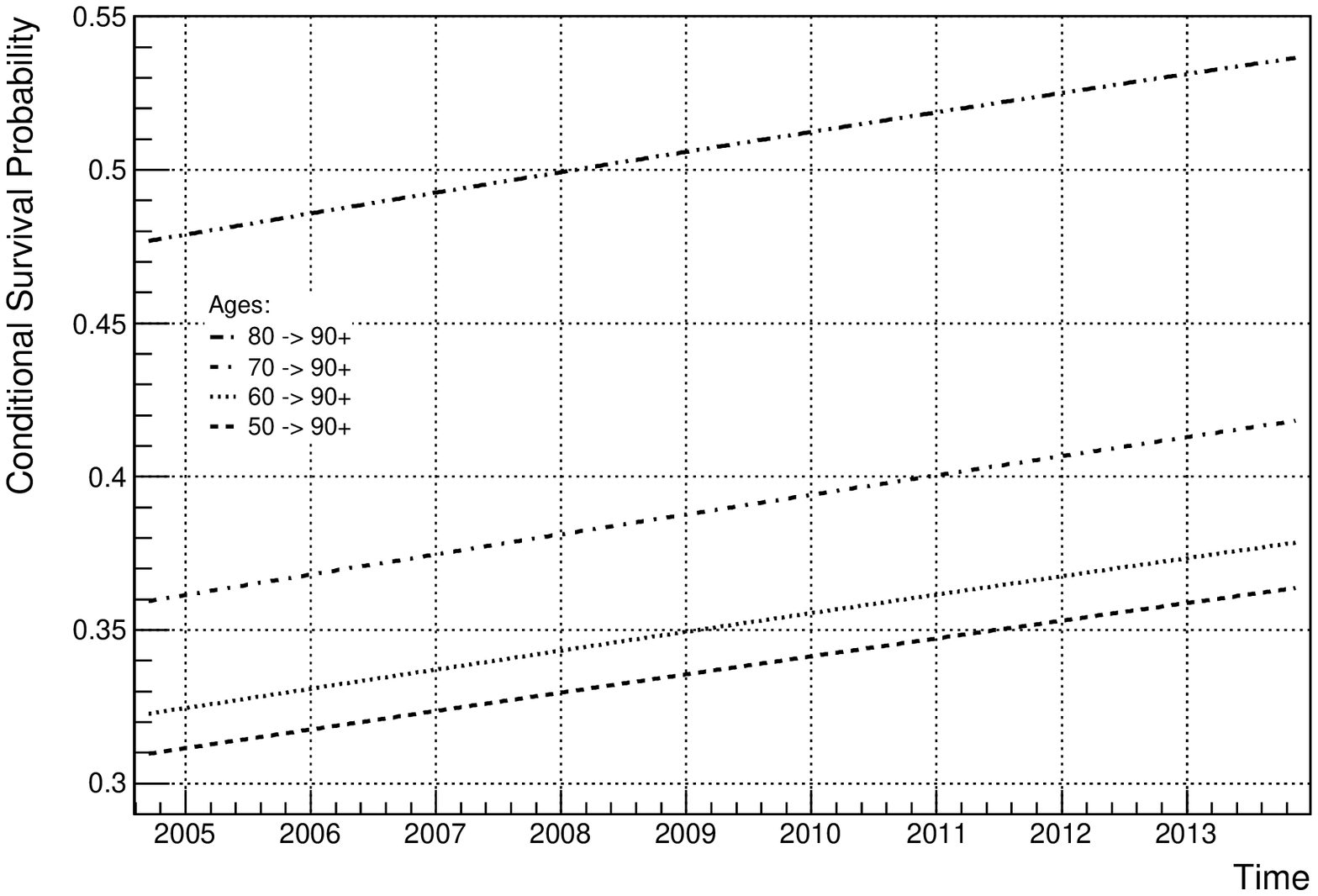}} 
  \caption{Implied Survival Probabilities (ISPs) derived from the GM parameters of the (curved) CIR model for MALES.}
  \label{FIG:MSBCIR} 
\end{figure}

Moving on to our main findings, Figure (\ref{FIG:FSBCIR}) {\bf [placed here]} displays ISPs from substituting the implied female mortality parameters given in Table (\ref{TAB:PARSMORT}) into the Gompertz-Makeham model. Figure (\ref{FIG:MSBCIR}) {\bf [placed here]} displays analogous results for males. These curves tell us how long the market thinks a typical annuitant (female and male) will live in terms of ISPs in any given week, based on the CIR term structure. Note that both indicate longevity improvements over time. For more concrete numbers, Table (\ref{TAB:ICSB}) {\bf [placed here]} shows the ISPs to the age of 90 beginning from ages 55, 65 and 75 at selected dates. Notice the increase in the ISP from September 2004 to November 2013. For example, in September 2004 annuity prices implied that a 75 year-old male had a $40.1\%$ probability of surviving to age 90. By November 2013 the implied value had increased to $46.1\%$. We also report the projected survival probabilities for the year 2020 assuming the current trend continues. According to our model, in the year 2020 75 year-old males will have a 49.7\% chance of living to the age of 90 while females of the same age will have a 54.3\% chance of attaining age 90. Table (\ref{TAB:ILE}) {\bf [placed here]} repeats the corresponding calculations for ILEs at ages 55, 65 and 75. For example, in September 2004 the ILE for a 75 year-old male was $13.09$ years. By November 2013, it had increased to 14.28 years, an increase of 6.78 weeks per year. We compare our market-based rates of increase with demographic-based rates in Table (\ref{TAB:HMD}) {\bf [placed here]}, and find roughly consistent results. The demographic numbers come from regressing mean lifetimes on date of death (1970--2010) from the period life tables in the Human Mortality Database, representing the overall U.S.A. population (for comment on alternative comparators, see the appendix). 

Last but not least, Table (\ref{TAB:APRICES}) {\bf [placed here]} gets to our main finding about annuity pricing. If interest rates had remained at September 2004 levels, annuity factors would have increased (and payouts would have declined) because longevity had improved. For example, a 65 year-old male would have paid $\$13.15$ for a dollar of lifetime income in 2013, compared to $\$12.73$ in 2004 (that is $3.3\%$ more expensive for a pension annuity) \emph{even} if interest rates had remained at 2004 levels\footnote{Note: we are puzzled by the age 55 female results which might be due to numerical instability and/or limitation of the GM law for young ages.}.

\section{Conclusion}
\label{SECconclusions}

\subsection{Summary of Results}
During the last decade life annuity payout rates in the U.S. have fluctuated from year to year but have overall have tumbled rather considerably. For example, in late October 2008 a 65 year-old male could have obtained an income of \$680 per month for life (guaranteed for 10 years), based on a \$100,000 premium\footnote{As we explained in the data section of the paper, this number is an average based on quotes offered by various U.S. insurance companies.}. By the end of the second half of the year 2013 (mid November 2013), the equivalent income generated by the same initial premium was a mere \$550, which is a reduction of 19\% in generated cash-flow, in nominal terms. Needless to say, in real terms the decline was even more pronounced. The same ``nest egg'' generated less income and a safe retirement has become more expensive in the last decade.

The conventional explanation for the declining trend in life annuity payouts has been the equivalent movement in U.S. interest rates during this period. Long-term interest rates went from 5\% to as high as 8\% to as low as 3\% over the course of the decade. But alas, the term structure of interest rates is only part of the story. A secondary and equally critical component in the decline of life payout rates has been the increase in the expected longevity of annuitants during this same 10-year period. An expectation of longer-lived annuitants implies that payments must last for longer which \emph{ceteris paribus} results in lower cash flows to the new generation of retirees. This is basic arithmetic.

Non-specialists mistakenly believe that changes to ``mortality tables'' underlying insurance pricing are uncovered with new studies and then implemented infrequently, perhaps once every few decades, and at the prodding of insurance regulators or actuarial associations. This paper and our study suggest that this is not the case, at least as it relates to market prices. We detect small and subtle changes from week to week. 

Methodologically we decompose the weekly changes in annuity payouts into two distinct components: (i.) the time-value-of-money interest rate changes, and (ii.) the expected biological or demographic changes. This then enables us to determine the relative importance of both components in explaining the decline in annuity payouts. Thus, for example, our methodology enables us to conclude that in early 2004 ``the market'' expected a typical 75-year old male annuitant to live to the age of 88 plus 1 month. By the end of the year 2013 the same market price implied a life expectancy of 89 years plus 3 months, which is an increase of approximately 1.5 months per year during the decade. Moreover, if this same trend continues, then market prices in the year 2020 are projected to imply a life expectancy of over 90 years for a 75-year old male who buys a life annuity in the year 2020 irrespective of where interest rates are at the time.

Tables (\ref{TAB:ICSB}) and (\ref{TAB:ILE}) (displaying ISPs and ILEs respectively) show the corresponding values for other ages. The main message is the same. The time-series of life annuity prices imply an increasing life expectancy over time. And, while our results are dependent on the assumed linear change in parameters driving the force of mortality, we soundly reject the hypothesis of no change at all. Our paper (and the detailed technical appendix) can also be viewed as contributing a methodology or algorithm on how to ``let the data speak'' and extract longevity expectations from a time-series of annuity prices. This same process could be applied to different markets and countries around the world to compare market-based longevity expectations without relying on demographic or actuaries tables.

Of course, we could have obtained the same sort of ``people are living longer'' insight by surveying the pricing actuaries within the insurance companies -- and in particular asking them about the longevity assumptions embedded in their changing mortality tables -- but we believe there is merit in understand what conclusions result from going directly to market prices. Our methodology and approach further confirms their independent statistical and biological projections. At the very least our study provides a good example of how the CIR model can be used to control for a latent interest rate process, when solving for the implied drivers of longevity. 

It is worth noting that the increase in longevity expectations for males over time has exceeded the increase in expectations for females. In other words, annuity prices are telling us that although (i.) male mortality rates are obviously higher than female mortality rates, and (ii.) the expectation is that both will continue to decline over time, markets are indicating that the improvement will be more pronounced for males. Of course, only time will tell whether -- like the futures market for frozen orange juice predicting the weather in Florida -- markets will be proven correct. 

\subsection{Practical Implications}

We envision two groups of practitioners who might be interested in these results and methodology, both within the pension and retirement arena. First, there is a growing market for pension {\em buy-outs and buy-ins} where DB plan sponsors reduce (or freeze or transfer) their liabilities using bulk annuity and longevity insurance arrangements. These deals are obviously predicated on certain mortality and longevity assumptions and the insurance company actuaries usually ``drive" this process. So, our methodology (as well as results) could be used by financial analysts (i.e. non actuaries) as an independent source to imply what various prices are saying about longevity assumptions. Perhaps, implied survival probabilities and life expectancies -- from such bulk arrangements -- can be compared against market annuity prices as a way of benchmarking prices \footnote{See Ezra (2007) for a discussion of this and related pension issues.}

A second (and perhaps larger) audience for this work is wealth management. A financial adviser or wealth manager must provide guidance to clients on more than dynamic {\em asset} allocation, or the mix of stocks and bonds in a diversified portfolio over time. As the client approaches retirement they need advice on optimal {\em product} allocation, that is the mix between pension annuity instruments and conventional non-insurance solutions. They also require guidance on optimal age-dependent spending and draw-down rates, which is another topic receiving more attention in the literature\footnote{See: Scott and Watson (2013)}. 

All of these retirement decisions are tightly connected and intertwined with longevity expectations. A client that is expected to live longer -- all else being equal -- should retire later, build a larger ``nest egg'', spend less in retirement and insure (more) against longevity risk. These are the normative implications of economic life-cycle theory, as well as common sense, for that matter\footnote{See for example the collection of articles by Horan (2009), and specifically chapter \#4 by Bodie, Treussard and Willen or chapter \#5 by Kotlikoff}.

More to the point, the advice that one often hears (perhaps anecdotally) is that annuitization or the purchase of an irreversible life annuity should be delayed or perhaps avoided entirely during periods of low interest rates. Others have argued for the existence of an option value in waiting to annuitize. And, while our intent here isn't to delve into the dynamic optimization of age-dependent annuitization policies, we end this paper with a note of caution. Namely, those who delay annuitization – for example from a retirement age of 68 to a draw-down age 73 – are swimming against the longevity extension tide. So yes, given the current interest rate and economic environment, a financial adviser might be reasonably confident that nominal interest rates will increase over the next 5 years, thus increasing annuity payouts. However, at the same time, we can only speculate whether the projected interest rate gain will outweigh the lost mortality credits and the more subtle longevity drift we have measured. In sum, annuity purchase procrastination might prove beneficial to health but hazardous to wealth.

\clearpage

\section{Technical Appendix}

\subsection{Models and Assumptions}
Underlying the methodology of this paper is the Gompertz-Makeham law of mortality given by 
\begin{equation}
  \lambda(x) = \lambda_0 + \fr{1}{b} \; e^{(x-m)/b}.
  \label{EQ:GML} 
\end{equation}
where the constants $\lambda_0$, $m$ and $b$ are explained earlier in the paper. The is a static expression that depends on age but does not depend on calendar time. 

One of the many ways to account for the changing longevity (parameters) over time is to make the modal and dispersion parameters $m$ and $b$ (used to price an annuity) depend on the week $z$ the annuity is purchased. So, to capture the changes in longevity the following first order perturbative expansion was deemed to be sufficient, considering that $m$ and $b$ change slowly with time:
\begin{gather}
  m(z)= m_0 + m_1 \, (z - z_0)  \label{EQ:MT},\\
  b(z) = b_0 + b_1 \, (z - z_0) \label{EQ:BT}.
\end{gather}
In the above, $m_1$ and $b_1$ are additional parameters responsible for incremental changes in longevity, $z$ is the purchase-date variable and $z_0$ is some reference date. The third parameter $\lambda_0$ in the Gompertz-Makeham (GM) model  represents ``accidental death'', and could in principle also vary over time. However, given the limitations of the data (just under $10$ years in length) it seems reasonable to focus only on variations  in the key actuarial parameters. For convenience we normalize so $z_0 = 0$ (i.e. we measure from the beginning of the data series) and choose units so the slopes $m_1$ and $b_1$ represent rates of change per week, to reflect our weekly data set. 

Swapping the ``no improvement in mortality'' constants for the linear function, the conditional probability of survival, from age $x$ to at least age $x+t$, is now denoted by:
\begin{equation}
  p(x, t\; | z) = e^{-\int_0^t \; \lambda(s+x\,| z) \; ds} 
  \label{EQ:MIRM} 
\end{equation}
where the $\lambda(s+x\,|z)$ is the purchase-date-dependent instantaneous rate of mortality. For instance, $p(65, 20, 17)$ is the conditional probability of survival to age $85$ for an individual purchasing an annuity at age 65, in the 17th week of our data series.

Equation (\ref{EQ:MIRM}) is quite general and should hold for any modification of the constants in the Gompertz-Makeham law of mortality or even other forms of mortality rate $\lambda(s+x| z)$. In the present (GM) setting, the integral in the exponential reduces to the following expression for the conditional probability of survival 
\begin{equation}
  p(x, t\, |z) = \exp\ds\left\{ - \lambda_0 \, t + \, e^{\frac{x-m(z)}{b(z)}} \left(1 - e^{\frac{t}{b(z)}}\right) \right\}.
  \label{EQ:MGMP} 
\end{equation}
Correspondingly, the remaining life expectancy $E[T_x \, |z]$ of an individual at age $x$ is also a function of purchase date $z$. It can be derived from equation (\ref{EQ:MGMP}), as shown in Milevsky (2006) or the standard actuarial textbooks such as Promislow (2011), and reduces to a convenient analytic expression $E[T_x \, |z] = \bar{a}(x, T, r = 0)$ as given by equation (\ref{GAPMformula}) with $m$ and $b$ replaced by $m(z)$ and $b(z)$. This expectation will change over time for two reasons: (i.) as the individual ages, and (ii.) as longevity improves. So, $E[T_x \, |z+1]>E[T_x \, |z] $, and $ E[T_{x+1} \, |z+1] > E[T_x \, |z] -1.$ 

In the actuarial and insurance literature there is an effect known as \emph{selection}; the population that selects by purchasing an annuity in week $z$ has different longevity parameters (hence different life expectancy) from the population that selects\footnote{Often, the act of purchasing an annuity signals greater health than average.} in week $z^\prime$. Appropriately, we don't assume any form of equivalence between $E[T_x \, |z]$ and $E[T_x \, |z^\prime]$ when $z\neq z^\prime$, and this distinction is important in interpreting our results. In particular, our model differs from those of Lee and Carter (1992) or Cairns, Blake and Dowd (2006), that preserve parameter consistency but forgo the simplicity of GM-pricing. 

Our model groups together individuals who purchase an annuity in a given year. This differs from the approach of actuarial period data (which groups individuals by year of death) or cohort data (grouping by year of birth). Actuaries also distinguish between the overall population and the (healthier) sub-population that purchases annuities. These distinctions strongly affect mean life expectancies, but should have less impact on rates of change, which is what we see in Table (\ref{TAB:HMD}). The demographic input to the latter is period data representing the overall U.S.A. population. A similar comparison could have been attempted based on the annuitant population, eg. using versions of the Society of Actuaries annuity tables from different years (though this would be complicated by the mortality improvement and loading assumptions incorporated into the latter). Alternatively, one could compare using cohort data, though to carry that out one would typically have to project future mortality using statistical models such as those of Lee and Carter (1992) or Cairns, Blake and Dowd (2006). See Pitacco et al (2009) for a discussion of these issues.

Moving back to annuity prices themselves, the basic annuity contract can be characterized by the starting age of the annuitant $x$ and the length of the guaranteed period $T$. The expression for the annuity factor can be derived from general arguments invoking the law of large numbers (to diversify away mortality risk) leaving deterministic cash-flows. It is the sum of the guaranteed and life contingent contributions. Following the convention in the literature, we may express the continuously-paid annuity factor using zero-coupon bonds:
\begin{equation}
  \bar{a}(x, T\, | z) = \int_0^T \, B(s) \, ds + \int_T^\infty \, B(s) \, p(x, s\, |z) \, ds,
  \label{EQ:MODA} 
\end{equation}
where $B(s)$ is the market price of a zero-coupon bond maturing at time $s$. This equation is valid regardless of the specific term structure model. In the flat term structure model, $B(s)=e^{-s\,r_z}$ and we obtain that $\bar{a}(x, T\, | z)= \bar{a}(x, T, r_z)$ where the latter is given in \eqref{GAPMformula} with, again, $m$ and $b$ replaced by $m(z)$ and $b(z)$. Note that $m(z)$ and $b(z)$ are always calculated on the day when the annuity factor is priced, i.e. they are constants for the purpose of calculating annuity factors. Annuity factors can be calculated once we have values of the parameters $r_z$, $\lambda_0$, $b_0$, $b_1$, $m_0$ and $m_1$.

To obtain a more realistic model of annuity valuation, we opted to use the one-factor CIR stochastic interest rate model, originally proposed in Cox, Ingersoll and Ross (1985). According to Kahn (2005) a one-factor model is able to explain up to $82\%$ of the interest rate variations in the US bond market. This enhancement -- from flat to curved -- provides an adequate way to describe an effective yield curve which insurance companies hypothetically use to price and value their annuity contracts. Under CIR, the instantaneous interest rate $R_t$, under the risk-neutral probability measure, evolves according to the following stochastic differential equation
\begin{equation}
  dR_t= (\alpha - R_t \, \beta) dt + \sigma \sqrt{R_t} \, dW_t,
  \label{EQ:CIR} 
\end{equation}
where $\alpha$, $\beta$ and $\sigma$ are all non-negative constants, $W_t$ is a one-dimensional Brownian motion, and the (week $z$) initial condition is $R_0=r_z$. The interest rate $R_t$ is a non-negative and mean-reverting process. Bond prices $B(s)$ are available for CIR in closed form -- see for example Shreve (2010). We acknowledge that the three global CIR parameters could also have been allowed to evolve in time, as in the Hull-White model, but our paper does not consider such issues. Even though bond prices are known, the two integrals in the expression for $\bar{a}(x, T\, |z)$ are not solvable analytically. Thus we resort to numerical integration to obtain annuity factors as functions of the GM parameters $\lambda_0$, $b_0$, $b_1$, $m_0$ and $m_1$, the CIR parameters $\alpha$, $\beta$, $\sigma$ and the weekly spot rate $r_z$.

A full model for annuity prices would include not just term structure (as above) but also loading factors. We have not done this because our data set spans only 10 years, and we already find significant redundancy between model parameters, which leads to high variances for certain estimates (eg for the CIR parameters). Adding a loading function would exacerbate this problem. To illustrate the nature of the redundancy, imagine a perfect fit to the data using the existing model. If we then modify the parameters $b_1$ and $m_1$ in an arbitrary way, we could subsequently restore the perfect fit by introducing a time-varying and age-dependent loading factor that precisely compensates for the change. Thus any estimation of loading parameters would depend strongly on the particular functional form assumed, and could be compensated for up to first order by varying the existing model parameters. We leave market price-based estimation of loadings question to the future, once richer data is available.

\subsection{Numerics and Methodology}
We reverse-engineer the problem of annuity valuation and seek the best values of the model parameters $(5+3+\text{weekly spot rates})$ from the data provided by CANNEX-QWeMA. We use the Levenberg-Marquardt (LM) nonlinear least squares algorithm for that purpose. This algorithm is described in Mor\"e (1978). It is a stepping method and requires the knowledge of the derivatives of the expression with respect to the desired parameters and initial guesses for the parameters themselves. In the case of the flat model, it is straightforward to work out these derivatives semi-analytically. For the CIR model, it is still feasible to calculate the derivatives semi-analytically, however these expressions are unreasonably lengthy and computationally extensive. We opted to use the three- and five-point difference methods to find derivatives. Essentially, they produce similar numerical accuracy but work much faster. 

The code performing the nonlinear least squares fitting was written in Python $2.7.3$ using the python bindings PyGSL $0.9.6$ for GSL $1.16$, which has an implementation of the LM algorithm and is well supported by its developers. In addition, among many other things, GSL provides numerical integration routines which we used to calculate the annuity factor of equation (\ref{EQ:MODA}) and its derivatives. All aforementioned software is open-source. The actual numerical computations were performed on the SHARCNET computing clusters and wherever possible the code was designed to run on multiple processors to speed up computation times. 
 
The female and male data was fitted separately resulting in two sets of longevity parameters and two nearly identical $r_z$-curves. Given the implausibility of males and females being offered or charged different interest rates (obviously prices are different), all $r_z$-curves and yields displayed and reported in this paper are averaged. The minor discrepancy between the two $r_z$-curves could be partially attributed to numerical noise. Another more likely explanation is the presence of a \emph{coupling effect} between the interest rate and longevity parameters. The coupling effect is dominant in the flat model because the LM algorithm attempts to compensate for the lack of a term structure by assigning the longevity parameters values different from those if a term structure was actually incorporated. 

In principle, the LM algorithm produces variances on the estimated parameter values. However, in our fitting methodology the Jacobian matrix $J$ is large and mostly populated with zeros because we treat every week with its unique parameter $r_z$. Consequently, $J$ is not of full rank, i.e. the columns of $J$ are almost linearly dependent. Therefore the variance-covariance matrix $(J^T\,J)^{-1}$ is close to being singular, as the numerical results showed, and the diagonal entries cannot be deemed as meaningful measures of error. Nonetheless, we possess a high degree of confidence that our results are valid for three reasons: (i.) it was verified that slightly perturbing the initial guesses of the parameters leads to similar results, (ii.) the fact that the parameters $b_0$ and $m_0$ are in agreement with actuarial projections of Lee and Carter (1992), (iii.) and the fact that the $r_z$-curves and yields are reasonable representation to the behaviour of the interest rates in the considered time period. This is our ``sanity'' check. 

In the curved (CIR) model, the interest rate and the longevity parameters must be non-negative as dictated by the model assumptions. Although, there are only a few (3) more parameters as compared to the flat model, it was disproportionally more difficult to establish their values. Moreover, the LM algorithm was not always able to find a set of estimated parameters with all positive values on initial runs. To circumvent this difficulty we restricted parameters to be positive by squaring them. However, this trick drastically slowed down convergence and lengthened computation times. The expressions (not written out in this paper) for the half-life, long term mean and variance of the interest rate were pertinent for selection of initial guesses as well as judging the reasonableness of the estimated parameters. Numerical computation was more extensive in the CIR model, where the LM algorithm had to go through hundreds of iterations and was computing individual runs for 3-5 days. 

\pagebreak

~\newline 

\clearpage

\begin{table}[!h]
  \begin{tabular}{ | c || c | c | c | c | c |}
    \hline                        
                &  $\lambda_0$     & $b_0$           & $b_1$     & $m_0$            & $m_1$ \\ \hline\hline
    female      & $1.622\pow{-3}$  & $9.058$  & $4.716\pow{-3}$  & $92.60$  & $1.044\pow{-3}$ \\\hline
    male        & $4.578\pow{-4}$  & $10.84$  & $4.268\pow{-3}$  & $88.47$  & $2.322\pow{-3}$ \\\hline
  \end{tabular}
    \vspace{0.15in}
    \caption{Point estimates for GM parameters under the flat term structure model. Parameters $b_1$ and $m_1$ measure rates of change per week.}
  \label{TAB:PARSFLAT} 
\end{table}

\clearpage

\begin{table}[!h]
  \begin{tabular}{ | c || c | c | c | c |}
    \hline                        
                &    n  &  $F$   & $F^{\left[p - g, n - k \right]}_{\left[\beta = 0.05 \right]}$ 
    &  $F^{\left[p -g, n - k \right]}_{\left[\beta = 0.01 \right]}$ \\ \hline\hline
    female      & 18756 &  2491.21  & 2.99622 & 4.60633  \\ \hline
    male        & 19100 &  4119.40  & 2.99621 & 4.60631  \\ \hline
  \end{tabular}
    \vspace{0.15in}
  \caption{Summary of the partial $F$-statistic test for the flat term structure model. The degrees of freedom constants are $p=484$, $g=482$ and $k=482$. The number of observations $n$ is $18,756 \,(19,100)$ for females (males). }
  \label{TAB:FTEST} 
\end{table}

\clearpage

\clearpage

\begin{table}[h]
  \begin{tabular}{ | c || c | c | c |}
    \hline                        
                & $\alpha$         & $\beta$        & $\sigma$         \\ \hline\hline
    female      & $1.804\pow{-3}$  & $7.210\pow{-3}$ & $4.093\pow{-2}$  \\ \hline
    male        & $1.731\pow{-3}$  & $3.731\pow{-3}$ & $4.341\pow{-2}$  \\ \hline
  \end{tabular}
    \vspace{0.15in}
  \caption{Point estimates for CIR parameters under the model with CIR (curved) term structure.}
  \label{TAB:PARSCIR} 
\end{table}

\clearpage

\begin{table}[h]
  \begin{tabular}{ | c || c | c | c | c | c |}
    \hline
                & $\lambda_0$      &  $b_0$   & $b_1$           & $m_0$           & $m_1$    \\ \hline\hline
    female      & $6.724\pow{-10}$ &  $9.174$ & $1.467\pow{-3}$ & $91.68$ & $8.201\pow{-4}$  \\ \hline
    male        & $2.376\pow{-10}$ &  $10.37$ & $1.127\pow{-3}$ & $88.13$ & $3.061\pow{-3}$  \\ \hline
  \end{tabular}
    \vspace{0.15in}
  \caption{Point estimates for GM parameters under the model with CIR (curved) term structure.}
  \label{TAB:PARSMORT} 
\end{table}

\clearpage

\clearpage

\clearpage

\clearpage

\clearpage

\begin{table}[p]
  \begin{tabular}{ | c || c | c | c | c | c | c |}
    \hline                        
    \multirow{2}{*}{Date}   & \multicolumn{2}{c |}{55} &  \multicolumn{2}{c |}{65}  & \multicolumn{2}{ c|}{75} \\ \hhline{~------}
                             &  Female   &  Male    & Female   & Male     & Female  &  Male   \\ \hline\hline
    September 2004           &  0.443    &  0.315   &  0.460   & 0.336    &  0.512  &  0.401 \\\hline 
    November 2013            &  0.455    &  0.369   &  0.474   & 0.393    &  0.531  &  0.461 \\ \hline 
    January 2020 (projected) &  0.463    &  0.404   &  0.484   & 0.423    &  0.543  &  0.497 \\ \hline 
  \end{tabular}
    \vspace{0.15in}
  \caption{Implied survival probabilities (ISPs) to age 90, under the CIR (curved) model, for individuals currently aged 55, 65, and 75.}
  \label{TAB:ICSB} 
\end{table}

\clearpage

\begin{table}[p]
  \begin{tabular}{ | c || c | c | c | c | c | c |}
    \hline                        
    \multirow{2}{*}{Date} & \multicolumn{2}{c |}{55} &  \multicolumn{2}{c |}{65}  & \multicolumn{2}{ c|}{75} \\ \hhline{~------}
                             &  Female  &   Male   & Female    &  Male    &  Female   & Male  \\ \hline\hline
    September 2004           &  32.13   &   28.72  &  23.10    &  20.30   &  15.08    & 13.09 \\\hline 
    November 2013            &  32.36   &   29.98  &  23.48    &  21.56   &  15.61    & 14.28 \\ \hline 
    January 2020 (projected) &  32.51   &   30.83  &  23.72    &  22.41   &  15.94    & 15.09 \\ \hline 
  \end{tabular}
    \vspace{0.15in}
  \caption{Implied life expectancies (ILEs) at ages 55, 65 and 75, under the CIR (curved) model.}
  \label{TAB:ILE} 
\end{table}

\clearpage

\begin{table}[p]
  \begin{tabular}{ | c || c | c | c | c | c | c |}
    \hline                        
    \multirow{2}{*}{Methodology} & \multicolumn{3}{c |}{Female} &  \multicolumn{3}{c |}{Male} \\ \hhline{~------}
                             &  55  &   65   & 75    &  55    &  65   & 75  \\ \hline\hline
    market price           &  1.31   &   2.16    &    3.02    &  7.18  &  7.18  & 6.78 \\\hline 
    demographic            &  4.44   &   3.61   &   2.64   &  7.78  &   5.98  &   3.52 \\ \hline 
  \end{tabular}
    \vspace{0.15in}
  \caption{Weeks per year improvement in life expectancies under the CIR (curved) model at ages 55, 65 and 75.}
  \label{TAB:HMD} 
\end{table}

\clearpage

\begin{table}[p]
  \begin{tabular}{ | c || c | c | c | c | c | c |}
    \hline                        
    \multirow{2}{*}{Date} & \multicolumn{2}{c |}{55} &  \multicolumn{2}{c |}{65}  & \multicolumn{2}{ c|}{75} \\ \hhline{~------}
                             &  Female    & Male       & Female    &  Male    &  Female  &  Male  \\ \hline\hline
    September 2004           &   16.42    &   15.44    &  13.87    &  12.73   &  10.56   &  9.47  \\ \hline 
    November 2013            &   16.39    &   15.71    &  13.93    &  13.15   &  10.76   &  10.05 \\ \hline 
    January 2020 (projected) &   16.38    &   15.88    &  13.97    &  13.42   &  10.89   &  10.42 \\ \hline 
  \end{tabular}
  \vspace{0.15in}
  \caption{Annuity prices (in \$) under the CIR (curved) model. These are calculated at selected dates, using the SAME spot interest rate as in September 2004, so price changes represent only the evolution of mortality.}
  \label{TAB:APRICES} 
\end{table}

\end{document}